\documentstyle[aps,prb,preprint]{revtex} 

\begin{document}  \preprint{\today} \draft
\title{Thermodynamic Properties of 1d Heisenberg Model.\\
Magnetic Moment of Spinon.}

\author{W. McRae and O. P. Sushkov\cite{Budker}}

\address{School of Physics, The University of New South Wales,
Sydney 2052, Australia}
\maketitle

\begin{abstract}
The one dimensional spin 1/2 Heisenberg antiferromagnet is considered using a simple quasiparticle picture - an interacting Fermi gas of kinks. Using this picture the low temperature heat capacity, and the magnetic susceptibility with logarithmic field and temperature corrections are derived. The results obtained are in agreement with previous computations and conformal field theory calculations. It is hypothesised that the magnetic moment of the kink $g_{kink}=\sqrt{2}g_{electron}$.
\end{abstract}

\vspace{0.5cm}



\narrowtext 
\section{Introduction}
We consider the 1d Heisenberg model with Hamiltonian
\begin{equation}
\label{H}
H=\sum_i{\bf S}_i{\bf S}_{i+1} +h\sum_i S_{iz}
\end{equation}
Here $S=1/2$ is spin localized at site $i$. The antiferromagnetic exchange is set
equal to unity: $J=1$. $h$ is an external magnetic field. This model has a very
special place in theoretical physics. The first major advance was made in the famous 
work of Bethe\cite{Bethe}, and the model has been studied in numerous subsequent works. 
For a review see the book of Mattis\cite{Mattis}. The most remarkable property of 
this model is its gapless spectrum of excitations  which is without spontaneously broken 
symmetry. Faddeev and Takhtajan\cite{Faddeev} found the structure of excitations at zero 
magnetic field. They demonstrated that the elementary excitations are topological kinks 
of spin 1/2 (spinons) with dispersion relation 
\begin{equation}
\label{eps1}
\epsilon(k)=v|\sin k|, \ \ \ \ v=\pi/2, \ \ \ \ \pi/2 \le k \le \pi/2.
\end{equation}
As usual we set lattice spacing equal to unity.

The heat capacity of the Heisenberg model was first calculated numerically by Bonner 
and Fisher\cite{Bonner}. Their result for low temperature was $C \approx 0.7 T$. 
Later Affleck\cite{Affleck}, using conformal field theory, derived that  $C = 2/3 T$. 
The zero field, zero temperature magnetic susceptibility was first calculated by 
Griffiths using the Bethe ansatz and numerics\cite{Griffiths}. His result agreed with 
$\chi =1/\pi^2$ within 5 decimal digits. Griffiths also pointed out the nonanalytic 
dependence of the susceptibility on the magnetic field. Quite recently Eggert, Affleck, 
and Takahashi\cite{Eggert}, using conformal field theory as well as numerical solution 
of the Bethe ansatz equations, calculated the temperature dependence of the zero field 
susceptibility: $\chi = 1/\pi^2 \left[ 1 +(2\ln7.7/T)^{-1} \right]$, at $T \ll 1$.

In the present paper we 1)  reproduce known results for heat capacity and
magnetic susceptibility using a simple quasiparticle approach, 2) find a previously 
unknown $\ln h$ correction,  3) comment on the value of the magnetic moment of a spinon.

\section{Interaction of the kinks. Effective dispersion}
Consider the scattering of two kinks of quasimomenta $k_1$ and $k_2$
resulting in two scattered kinks of quasimomenta $k_1^{\prime}$ and
$k_2^{\prime}$. Conservation of energy and momentum requires 
$k_1=k_1^{\prime}$, $k_2=k_2^{\prime}$. This means that the only change to
the wave function of the kink is an additional phase factor: $\exp(ikn) \to
S(k) \exp(ikn)$, $S(k)=\exp(i\phi)$. According to Faddeev and Takhtajan\cite{Faddeev}
$S(k)$ is given by
\begin{eqnarray}
\label{S}
S_t(k)&=&-i {{\Gamma(1-i\lambda/2) \Gamma(1/2+i\lambda/2)}\over
               {\Gamma(1+i\lambda/2) \Gamma(1/2-i\lambda/2)}}
 \approx \exp({i\over{2\lambda}})\nonumber\\
S_s(k)&=&{{\lambda-i}\over{\lambda+i}}S_t(k) \approx \exp({{-3i}\over{2\lambda}})\\
\lambda &=& {1\over{\pi}}\sinh^{-1}(\cot k) \approx \ln{2\over{|k|}},\nonumber
\end{eqnarray}
for triplet (total spin unity) and singlet (total spin zero) scattering. The right hand 
side of eqs. (\ref{S}) corresponds to the low energy limit: $k \ll 1$, $\lambda \gg 1$. 
Using perturbation theory we can relate the phase factor $S(k)$ to the matrix 
element of the interaction between the kinks
\begin{eqnarray}
 \label{HH}
 |k \rangle_{r \to \infty}^{\prime} &=&\left(|k\rangle +
\sum_p{{\langle p|H_{int}|k\rangle}\over
 {\epsilon_k - \epsilon_p +i0}} e^{ipr}\right)_{r \to \infty} =
e^{ikr}\left( 1-{{\langle k|H_{int}|k\rangle}\over{d\epsilon/dk}}\int{{e^{iqr}}
\over{(q-i0)}}
{{dq}\over{2\pi}}\right)=\\
&&e^{ikr}\left[ 1- i{{\langle k|H_{int}|k\rangle}\over v}\right]=e^{ikr}\exp
\left(-i{{\langle k|H_{int}|k\rangle}\over v}\right) \nonumber
 \end{eqnarray}
Comparison with eqs.(\ref{S}) yields the matrix elements of the triplet and
the singlet interaction. It is more convenient to work with states of fixed
projection of the kink's spin. We can easily find
\begin{equation}
\label{Hint}
\langle k_1 \uparrow , k_2 \uparrow |H_{int}| k_1 \uparrow , k_2 \uparrow \rangle
=-\langle k_1 \uparrow , k_2 \downarrow | H_{int}| k_1 \uparrow , k_2 \downarrow \rangle = -v^2 \left({1\over{\ln |2/k_1|}} + {1\over{\ln |2/k_2|}}\right).
\end{equation}
Combining eqs. (\ref{eps1}) and (\ref{Hint}) we find dispersion of the kink
$(k \ll 1)$ with spin up (down) in the presence of other kinks
\begin{equation}
\label{eps2}
\epsilon_{\pm}(k)=v|k| \mp {{v^2}\over{\ln |2/k|}}\left(n_+ -n_-\right) \mp gh.
\end{equation}
Here $n_+$ ($n_-$) is concentration of the spin up (down) kinks.
We have introduced also the interaction of the kink with the external magnetic field:
$\mp gh$. $g$  is the magnetic moment of the kink.

\section{Heat Capacity}
The number of kinks is not conserved, so the chemical potential is equal to zero:
$\mu =0$. The spin of the kink equals 1/2, therefore the kinks obey the usual 
Fermi-Dirac distribution: $f_{\pm}=1/\left(\exp(\epsilon_{\pm}/T)+1\right)$.
It is well known that in one dimension interactions can influence the statistics
of the quasiparticles. For example hard-core bosons are classified
as fermions\cite{Mattis}. Fractional statistics are also possible\cite{Haldane}.
However for the kinks in Heisenberg model, according to eq.(\ref{Hint})
the interaction vanishes in the low energy limit. Therefore there is no
reason for a deviation from simple Fermi statistics.

In the absence of a magnetic field the numbers of spin up and spin down
kinks are equal, so the interaction term disappears from the dispersion (\ref{eps2}).
The low temperature and zero magnetic field heat capacity equals
\begin{equation}
\label{C}
C={{\partial}\over{\partial T}} 2\int_{-\infty}^{+\infty} {{dk}\over{2\pi}}
{{v|k|}\over{\left[\exp(v|k|/T)+1\right]}}
= {2\over 3} T.
\end{equation}
This agrees with both the numerical result of Bonner and Fisher\cite{Bonner}
and the conformal field theory result of Affleck\cite{Affleck}.

\section{Magnetic Susceptibility}
The zero temperature, zero field magnetic susceptibility can be calculated by 
considering an ideal Fermi gas with dispersion $\epsilon_{\pm}=v|k| \mp gh$. 
Spin down states are empty, and spin up states are occupied up to quasimomentum
$k_{max}$ which is defined by the condition of zero chemical potential:
$\epsilon(k_{max})=\mu=0$. Magnetic  moment per unit length equals
\begin{equation}
\label{M}
M=gn_+=g \int_{-k_{max}}^{k_{max}}{{dk}\over{2\pi}}={{g^2h}\over{\pi v}},
\end{equation}
and magnetic susceptibility $\chi=dM/dh=g^2/(\pi v)$. To reproduce known
result $\chi=1/\pi^2$ (Refs.\cite{Griffiths,Eggert}) we need to set $g=1/\sqrt{2}$.
Thus the magnetic moment of the kink
\begin{equation}
\label{g}
g=g_{kink}=\sqrt{2}g_{electron}.
\end{equation}
We will comment later on this relation. We now derive $\ln T$ and $\ln h$
corrections.

In the case $h \ll T \ll 1$, the typical value of $k$ is of the order of $T$.
Therefore  under the logarithm in eq.(\ref{eps2}), $k$ can be replaced by $T$:
$v^2/\ln|2/k| \to \alpha=v^2/\ln|2/T|$. The magnetic moment equals
\begin{equation}
\label{MT}
M=g(n_+-n_-)=2g\int_0^{\infty}{{dk}\over{2\pi}}\left(
f[vk-gh-\alpha(n_+-n_-)]-f[vk+gh+\alpha(n_+-n_-)]\right).
\end{equation}
Expanding the right hand side of this equation in powers of $\alpha$ and $h$ we
can easily find $n_+-n_-$ and the susceptibility
\begin{equation}
\label{chit}
\chi = {1\over{\pi^2}}\left(1+{1\over{2\ln T_0/T}}\right),
\end{equation}
where $T_0=2$. This result is almost the same as the result of 
Eggert {\it et al}\cite{Eggert}, but differs in the value of $T_0$. Eggert {\it et al}
chose $T_0 \approx 7.7$ to provide a best fit with numerical data calculated
from the Bethe ansatz. The renormalization of $T_0$ ($T_0=2 \to T_0 \approx 7.7$) 
effectively takes account of double logarithm correction ($\sim 1/\ln^21/T$).

Now we consider the $\ln h$ correction at zero temperature: $T\ll h \ll 1$.
Spin down states are empty and spin up states are occupied up to $k_{max}$,
given by the condition $\epsilon_+(k_{max})=0$. Due to eq.(\ref{eps2})
this gives
\begin{eqnarray}
\label{lnh}
&&vk_{max}-gh-{{v^2}\over{\ln|2/k_{max}|}}=0\\
&&n_+=\int_{-k_{max}}^{k_{max}}{{dk}\over{2\pi}}={{k_{max}}\over{\pi}}\nonumber
\end{eqnarray}
Solving these eqs. we find $n_+$ and the susceptibility
\begin{equation}
\label{chih}
\chi = {1\over{\pi^2}}\left(1+{1\over{2\ln h_0/h}}\right),
\end{equation}
where $h_0=2v/g \approx 4.4$. To fit the zero temperature magnetization curve 
calculated by Griffiths\cite{Griffiths} we need to set $h_0=17.4$. This means
that the renormalization of $h_0$ due to the double logarithm corrections
is by a factor of 4 -- exactly similar to the renormalization of $T_0$ in the case 
of temperature dependence (\ref{chit}).

\section{Conclusions}
In the present paper we haved used a quasiparticle approach to calculate
the heat capacity and the magnetic susceptibility including $\ln T$ and $\ln h$
corrections. The results agree with previous analytical and numerical
calculations. The most interesting point is probably the value of the magnetic
moment of kink: $g_{kink}$. To reproduce the known value of the zero temperature, 
zero field magnetic susceptibility we set $g_{kink}=\sqrt{2} g_{electron}$. Naively 
one would expect that $g_{kink}=g_{electron}$ because total spin ${\bf S}_{tot}=
\sum_i{\bf S}_i$ is conserved. The violation of this naive expectation is
probably an indication of some anomaly.

\section{Acknowledgments}
We are very grateful to J. Bellissard, V. Flambaum, V. Kotov, D. Khomskii,
M. Kuchiev, D. Mattis, M. Mostovoy, J. Oitmaa, G. Sawatzky and R. Singh for 
stimulating discussions.

\tighten

\end{document}